\documentclass[aps, prb, twocolumn, showpacs]{revtex4}
\usepackage{amsmath, amsfonts, amssymb}
\usepackage{bm}
\usepackage{hyperref}
\usepackage{graphicx}

\begin{document}

\def\Re {\mbox{Re}}
\def\Im {\mbox{Im}}
\newcommand{\avg}[1]{\langle#1\rangle}
\newcommand{\odiff}[2]{\frac{\di #1}{\di #2}}
\newcommand{\pdiff}[2]{\frac{\partial #1}{\partial #2}}
\newcommand{\di}{\mathrm{d}}
\newcommand{\ii}{i}
\newcommand{\norm}[1]{\left\| #1 \right\|}
\renewcommand{\vec}[1]{\mathbf{#1}}
\newcommand{\ket}[1]{|#1\rangle}
\newcommand{\bra}[1]{\langle#1|}
\newcommand{\pd}[2]{\langle#1|#2\rangle}
\newcommand{\tpd}[3]{\langle#1|#2|#3\rangle}
\renewcommand{\vr}{{\vec{r}}}
\newcommand{\vk}{{\mathbf{k}}}
\renewcommand{\ol}[1]{\overline{#1}}

\title{Non-adiabatic Effects in the Braiding of Non-Abelian Anyons in Topological Superconductors}
\author{Meng Cheng} \author{Victor Galitski} \author{S. Das Sarma}
\affiliation{Condensed Matter Theory Center, Department of Physics,
            University of Maryland, College Park, Maryland 20742, USA\\ }
\date{\today}

\begin{abstract}
	Qubits in topological quantum computation are built from non-Abelian anyons.  Adiabatic braiding of anyons is exploited as topologically protected logical gate operations. Thus, the adiabaticity upon which the notion of quantum statistics is defined, plays a fundamental role in defining the non-Abelian anyons. We study the non-adiabatic effects in braidings of Ising-type anyons, namely Majorana fermions in topological superconductors, using the formalism of time-dependent Bogoliubov-de Gennes equations. Using this formalism, we consider non-adiabatic corrections to non-Abelian statistics from: (1) tunneling splitting of anyons imposing an additional dynamical phase to the transformation of ground states; (2) transitions to excited states that are potentially destructive to non-Abelian statistics since the non-local fermion occupation can be spoiled by such processes. However, if the bound states are localized and being braided together with the anyons, non-Abelian statistics can be recovered once the definition of Majorana operators is appropriately generalized taking into account the fermion parity in these states. On the other hand, if the excited states are extended over the whole system and form a continuum, the notion of local fermion parity no longer holds. We then quantitatively characterize the errors introduced in this situation.
\end{abstract}
\maketitle

\section{Introduction}
Fragility of quantum coherence against external perturbations is one of the main obstacles in building a quantum computer. Maintaining coherence in the presence of environment is thus among the very first challenges in the realization of quantum computation. Topological Quantum Computation(TQC)~\cite{Kitaev_AP03,nayak_RevModPhys'08} provides an elegant solution to the problem: quantum information is encoded and manipulated in a non-local way such that weak local perturbation has essentially no influence on the computation procedure so the decoherence is eliminated at the hardware level. To achieve this goal, qubits are built out of non-Abelian anyons, which are exotic particles obeying non-Abelian statistics, and adiabatic braidings of the anyons generate quantum gates. The intrinsic protection of quantum coherence against environmental noise makes TQC a very appealing scheme for quantum computation. It is thus very desirable to understand the properties of non-Abelian anyons as well as to search for quantum phases of matter that host them, known as non-Abelian topological phases.

Quantum statistics describes how many-body wavefunctions transform under the exchange of any pair of particles. The unitary transformations form representations of the braid group.~\cite{Wilczek_book1990} In (2+1)-dimensions, non-Abelian statistics happens when there are degenerate ground states, and braidings(pair-wise exchanges) result in unitary rotations in the ground state subspace, corresponding to higher-dimensional representation of the braid group in two-dimensions. These braiding operations form the basic set of topologically protected gates to perform quantum computation. The simplest of such anyons, and probably the only one which can be potentially realized, is the so-called Ising-type anyons. It has a natural interpretation as Majorana fermions which are self-conjugate neutral fermionic excitations. Such excitations are found to exist in some topological superconducting systems such as two-dimensional spinless chiral p-wave superconductors. They are in one-to-one correspondence with non-degenerate zero-energy states in superconductors which exist in vortices in two-dimensions~\cite{Kopnin_PRB'91, volovik_JETP'99, read_prb'00, Ivanov_PRL'01, Gurarie_PRB'07} or domain walls in one-dimension~\cite{Kitaev_Majorana}. Non-Abelian statistics of Majorana fermions in both cases have been demonstrated explicitly~\cite{Ivanov_PRL'01, Alicea_NatPhys2011} and are also supported by rather general arguments based on the conservation of fermion parity~\cite{Sau_PRA2010}. For experimental realizations, it has long been conjectured that quasiholes in $\nu=5/2$ fractional quantum Hall state are Ising anyons~\cite{Moore_NPB91, Nayak_NPB96, Bonderson_PRB2011} and subsequently candidate materials such as Sr$_2$RuO$_4$ for chiral p-wave superconductivity were also discovered~\cite{Mackenzie_RevModPhys'03}. However, recent theoretical advances have provided a variety of different heterostructure systems, where spin-orbit coupled semiconductors~\cite{Sau_PRL10, Alicea_PRB10, Lutchyn_PRL2011, Oreg_PRL2010} or the surfaces of three-dimensional topological insulators~\cite{Fu_PRL08} combined with proximity-induced s-wave pairing are shown to successfully engineer spinless $p_x+ip_y$ superconductivity with non-Abelian Majorana anyons.

Mathematical definition of quantum statistics necessarily builds upon the concept of Berry phase of many-body wavefunctions. This implies that the adiabaticity of braiding is an essential ingredient for non-Abelian statistics, since the quantum state has to stay in the ground state manifold during the entire process of the braiding. In the real world, however, braidings are necessarily performed within a finite time interval, {\it i.e.}, they are always non-Adiabatic. As known from the adiabatic perturbation theory, Berry phase is the leading-order term in the adiabatic perturbative expansion.~\cite{Berry_1984, Rigolin_PRA2008, Rigolin_PRL2010}  Given the fundamental role played by adiabatic braiding in TQC, it is therefore important to understand quantitatively the higher order corrections arising from non-adiabatic evolution,

In this paper, we present a systematic study of the non-adiabatic corrections to the braiding of non-Abelian anyons and develop formalism to describe their dynamical aspects. In our treatment, braidings are considered as dynamical evolutions of the many-body system, essentially using the time-dependent Schr\"odinger equation of the BCS condensate whose solutions are derived from time-dependent Bogoliubov-de Gennes (BdG) equation. Generally, adiabaticity may break down in three different ways: (a) tunneling of non-Abelian anyons when there are multiples of them, which splits the degenerate ground state manifold and therefore introduces additional dynamical phases in the evolution; (b) transitions to excited bound states outside the Hilbert space of zero-energy states, in this case topologically protected braidings have to be defined within an enlarged Hilbert space; (c) transitions to the continuum of extended states which render the fermion parity in the low-energy Hilbert space ill-defined. These non-adiabatic effects are possible sources of errors for quantum gates in TQC. The main goal of this paper is to quantitatively address these effects and their implications on quantum computation. 

Our work is the first systematic attempt to study non-adiabaticity in the anyonic braiding of non-Abelian quantum systems.  Given that the braiding of non-Abelian anyons is the unitary gate operation~\cite{dassarma_prl'05} in topological quantum computation~\cite{nayak_RevModPhys'08}, understanding the dynamics of braiding as  developed in this work is one of the keys to understanding possible errors in topological quantum computation.  The other possible source of error in topological quantum computation is  the lifting of the ground state anyonic degeneracy due to inter-anyon tunneling, which we have studied elsewhere~\cite{Cheng_PRL09, Cheng_PRB2010b}.  Although we study the braiding non-adiabaticity in the specific context of the  topological chiral p-wave superconductors using the dynamical BdG equatons within the BCS theory, our work should be of general validity to all known topological quantum computation platforms, since all currently known non-Abelian anyonic platforms in nature are based on the $\mathbb{SU}(2)_2$ conformal field theory of Ising anyons, which are all isomorphic to the chiral p-wave topological superconductors~\cite{nayak_RevModPhys'08}.  As such, our work, with perhaps some minor modifications in the details, should apply to the fractional quantum Hall non-Abelian qubits~\cite{dassarma_prl'05},  real p-wave superconducting systems based on solids~\cite{SDS_PRB06} and quantum gases~\cite{tewari_prl'2007}, topological insulator-superconductor heterostructures~\cite{Fu_PRL08}, and semiconductor-superconductor sandwich structures~\cite{Sau_PRL10} and nanowires~\cite{Lutchyn_PRL2011, Lutchyn_PRL2011b}.  Our results are quite general and are independent, in principle, of the detailed methods for the anyonic braiding which could vary from system to system in details.

The reminder of the paper is organized as follows: in Sec. \ref{sec1} we review Majorana fermions in non-Abelian topological superconductors and their quantum statistics defined in terms of adiabatic evolution. In Sec. \ref{sec:bdg} we generalize BdG equation to describe time-evolution of BCS superconductors under a parametrically time-dependent Hamiltonian.  Then we study the three main non-adiabatic effects as outlined above using the approach of time-dependent BdG equation.

\section{Quantum Statistics of Majorana Fermions}
\label{sec1}
\subsection{Quantum Statistics and Adiabatic Evolution}
\label{sec:nonabelian}
We first briefly review how quantum statistics is formulated mathematically in terms of the adiabatic evolution of many-body wavefunctions, following a recent exposition in Ref. [\onlinecite{Bonderson_PRB2011}].  Consider the general many-body Hamiltonian $\hat{\mathcal{H}}[\vec{R}_1(t),\dots,\vec{R}_n(t)]$ where parameters $\{\vec{R}_i\}$ represent positions of quasiparticles. We assume the existence of well-defined, localized excitations which we call quasiparticles. At each moment $t$, there exists a subspace of instantaneous eigenstates of $\hat{\mathcal{H}}[\vec{R}_1(t),\dots,\vec{R}_n(t)]$ with degenerate energy eigenvalues. Instantaneous eigenstates in the subspace are labeled as $\ket{\alpha(t)}\equiv\ket{{\alpha}(\{\vec{R}_i(t)\})}$. We constraint our discussion in the ground state subspace with zero energy eigenvalue.

The adiabatic exchange of any two particles can be mathematically implemented by adiabatically changing the positions of two particles, say $\vec{R}_i$ and $\vec{R}_j$, in such a way that in the end they are interchanged. This means that 
\begin{equation}
\vec{R}_i(T)=\vec{R}_j(0), \vec{R}_j(T)=\vec{R}_i(0).
\end{equation}
According to the adiabatic theorem~\cite{bohm_qm}, it results in a unitary transformation within the subspace: if the system is initially in state $\ket{\psi(0)}$, then $\ket{\psi(T)}=\hat{U}\ket{\psi(0)}$. To determine $\hat{U}$, we first consider initial states $\ket{\psi(0)}=\ket{\alpha(\{\vec{R}_i(0)\})}$. Under this evolution the final state can be written as
\begin{equation}
	\ket{\psi_\alpha(T)}=\hat{U}_0\ket{\alpha(T)}.
	\label{eqn:adibevol1}
\end{equation}
Here the matrix $\hat{U}_0$ is the non-Abelian Berry phase:\cite{Berry_1984, Simon_PRL1983, Wilczek_PRL1984}
\begin{equation}
	\hat{U}_0=\mathcal{P}\exp\left( i\int_0^T \di t\,\hat{\mathcal{M}}(t) \right),
	\label{}
\end{equation}
where $\mathcal{P}$ denotes path ordering and matrix element of the Berry's connection $\hat{\mathcal{M}}$ is given by
\begin{equation}
	\hat{\mathcal{M}}_{\alpha\beta}(t)=i\pd{\alpha(t)}{\dot{\beta}(t)}.
	\label{}
\end{equation}

Although the exchange defines a cyclic trajectory in the parameter space of Hamiltonian, the final basis states can be different from the initial ones (e.g, $\ket{\alpha(\{\vec{R}_i\})}$ can be multivalued functions of $\vec{R}_i$, which is allowed if we are considering quasiparticles being collective excitations of many-body systems). The only requirement we impose is that the instantaneous eigenstates $\{\alpha(t)\}$ are continuous in $t$. Therefore, we have another matrix $\hat{B}$ defined as $\hat{B}_{\alpha\beta}\equiv \pd{\alpha(0)}{\beta(T)}$, relating $\{\alpha(T)\}$ to $\{\alpha(0)\}$: $\ket{\alpha(T)}=\hat{B}_{\alpha\beta}\ket{\beta(0)}$. Combining with \eqref{eqn:adibevol1}, we now have the expression for $\hat{U}$:
\begin{equation}
	\ket{\psi(T)}=\hat{U}_0\hat{B}\ket{\psi(0)}.
\end{equation}
Therefore 
\begin{equation}
\hat{U}=\hat{U}_0\hat{B}=\mathcal{P}\exp\left( i\int_0^T \di t\,\hat{\mathcal{M}}(t) \right)\hat{B}.
\end{equation}
In fact, the factorization of $\hat{U}$ into $\hat{U}_0$ and $\hat{B}$ is somewhat arbitrary and gauge-dependent. However, their combination $\hat{U}$ is gauge-independent provided that the time-evolution is cyclic in parameter space(positions of particles). The unitary transformation $\hat{U}$ defines the statistics of quasiparticles.

\subsection{Non-Abelian Majorana Fermions}
We now specialize to the non-Abelian statistics of Majorana fermions in topological superconductors, carefully treating the effect of Berry phases. We mainly use spinless superconducting fermions as examples of topological superconductors in both 1D and 2D since all known topological superconducting systems supporting non-Abelian excitations essentially stem from spinless chiral p-wave superconductors.\cite{read_prb'00, Alicea_PRB10, Fu_PRL08}

In the BCS mean-field description of superconductors, the Hamiltonian is particle-hole symmetric due to $\mathbb{U}(1)$ symmetry breaking. In terms of Nambu spinor $\hat{\Psi}(\vr)=(\hat{\psi}(\vr), \hat{\psi}^{\dag }(\vr))^T$, the BCS Hamiltonian is expressed as $\hat{\mathcal{H}}_\text{BCS}=\frac{1}{2}\int\di^2\vr \hat{\Psi}^\dag(\vr)H_\text{BdG}\hat{\Psi}(\vr)$. The Bogoliubov-de Gennes Hamiltonian $\bm{H}_\text{BdG}$ takes the following form~\cite{Stern_PRB'04,read_prb'00}
\begin{equation}
	\bm{H}_\text{BdG}=
	\begin{pmatrix}
		\bm{h} & \bm{\Delta}\\
		\bm{\Delta}^\dag & -\bm{h}^T
	\end{pmatrix},
	\label{eqn:hamiltonian}
\end{equation}
where $\bm{h}$ is the single-particle Hamiltonian [for spinless fermions it is simply $\bm{h}=\left(-\frac{1}{2m}\partial_\vr^2-\mu\right)\delta(\vr-\vr')$] and $\Delta$ is the gap operator.  The BCS Hamiltonian can be diagonalized by Bogoliubov transformation
\begin{equation}
\hat{\gamma}^\dag= \int \di^2\vec{r}\,\big[u(\vr)\hat{\psi}^\dag(\vec{r})+v(r)\hat{\psi}(\vec{r})\big].
\end{equation}
Here the wavefunctions $u(\vr)$ and $v(\vr)$ satisfy BdG equations:
\begin{equation}
\bm{H}_\text{BdG}
\begin{pmatrix}
	u(\vr)\\
	v(\vr)
\end{pmatrix}=E
\begin{pmatrix}
	u(\vr)\\
	v(\vr)
\end{pmatrix}.
\end{equation}
Throughout this work, we adopt the convention that operators which are hatted are those acting on many-body Fock states while bold ones denote matrices in ``lattice'' space.

The single-particle excitations $\hat{\gamma}$, known as Bogoliubov quasiparticles, are coherent superpositions of particles and holes. The particle-hole symmetry implies that the quasiparticle with eigenenergy $E$ and that with eigenenergy $-E$ are related by $\hat{\gamma}_{-E}=\hat{\gamma}_E^\dag$. Therefore, $E=0$ state corresponds to a Majorana fermion $\hat{\gamma}_0=\hat{\gamma}_0^\dag$~\cite{Gurarie_PRB'07}. The existence of such zero-energy excitations also implies a non-trivial degeneracy of ground states: when there are $2N$ such Majorana fermions, they combine pair-wisely into $N$ Dirac fermionic modes which can either be occupied or unoccupied, leading to $2^{N}$-fold degenerate ground states. The degeneracy is further reduced to $2^{N-1}$ by fermion parity~\cite{nayak_RevModPhys'08}.  Since these fermionic modes are intrinsically non-local, any local perturbation can not affect the non-local occupancy and thus the ground state degeneracy is topologically protected. This non-locality lies at the heart of the idea of topological qubits.  

We are mostly interested in Majorana zero-energy states that are bound states at certain point defects ({\it e.g.} vortices in 2D, domain walls in 1D). In fact, Majorana bound states are naturally hosted by defects because that zero-energy states  only appear when gap vanishes. Defects can be moved along with the Majorana fermions bound to them. Braidings of such Majorana fermions realizes very non-trivial non-Abelian statistics.

We now apply the general theory of quantum statistics as previously discussed in Sec. \ref{sec:nonabelian} to the case of Majorana fermions in topological superconductors. The simplest setting where non-trivial statistics can be seen is the adiabatic braiding of two spatially separated Majorana fermions $\hat{\gamma}_1$ and $\hat{\gamma}_2$. We denote the two bound state solutions of the BdG equation by $\Psi_{01}$ and $\Psi_{02}$. When $\vec{R}_1$ and $\vec{R}_2$ vary with time they become instantaneous zero-energy eigenstates of BdG Hamiltonian. We choose their phases in such a way that the explicit analytical continuation of BdG wavefunction leads to the following basis transformation under exchange~\cite{Ivanov_PRL'01} 
\begin{equation}
 \begin{gathered}
	\Psi_{01}(T) = s\Psi_{02}(0)\\
	\Psi_{02}(T) = -s\Psi_{01}(0),
\end{gathered}
  \label{}
\end{equation}
where $s=\pm 1$. The value of $s$ depends on the choice of wavefunctions and we choose the convention that $s=1$ throughout this work. In the case of Majorana fermions in vortices, the additional minus sign originates from branch cuts introduced to define the phase of wavefunctions. This transformation actually gives the $\hat{B}$ matrix in the general theory.  Equivalently in terms of quasiparticle operator, we have \begin{equation}
  \begin{gathered}
 	\hat{\gamma}_1\rightarrow \hat{\gamma}_2\\
    \hat{\gamma}_2\rightarrow -\hat{\gamma}_1.   
  \end{gathered}
	\label{eqn:ivanov}
\end{equation}

If we define the non-local fermionic mode $\hat{d}^\dag=\frac{1}{\sqrt{2}}(\hat{\gamma}_1+i\hat{\gamma}_2)$, the states with even and odd fermion parity are given by $\ket{\text{g}}$ and $\hat{d}^\dag \ket{\text{g}}$. Due to the conservation of fermion parity, the two states are never coupled. However, the non-Abelian statistics still manifest itself in the phase factor acquired by the two states after an adiabatic exchange. To see this, first we notice that under exchange, the analytical continuation(or basis transformation) gives the following transformation of the two states:
\begin{equation}
  \begin{gathered}
 \ket{\text{g}}\rightarrow e^{i\varphi}\ket{\text{g}}\\
 \hat{d}^\dag \ket{\text{g}}\rightarrow e^{i\frac{\pi}{2}}e^{i\varphi}\hat{d}^\dag\ket{\text{g}} .  
  \end{gathered}
  \label{eqn:nonabelian2}
\end{equation}
Here the $\frac{\pi}{2}$ phase difference is reminiscence of non-Abelian statistics.

So far we have obtained the basis transformation matrix $\hat{B}$. To know the full quantum statistics we also need to calculate the adiabatic evolution $\hat{U}_0$. We now show by explicit calculation that $\hat{U}_0\propto \hat{1}$ up to exponentially small corrections. This requires knowledge of Berry connection accompanying adiabatic evolutions of BCS states. Fortunately, for BCS superconductor the calculation of many-body Berry phase can be done analytically~\cite{stone_prb'06}.  The ground state $\ket{\text{g}}$ has the defining property that it is annihilated by all quasiparticle operators $\gamma_n$. All other states can be obtained by populating Bogoliubov quasiparticles on the ground state $\ket{\text{g}}$. Let us consider a state with $M$ quasiparticles $\ket{n_1,n_2,\dots,n_M}=\hat{\gamma}_{n_1}^\dag\hat{\gamma}_{n_2}^\dag\cdots\hat{\gamma}_{n_M}^\dag\ket{\text{g}}$. The Berry connection of this state then reads~\cite{stone_prb'06}
\begin{multline}
\tpd{n_1,\dots,n_M}{\partial}{n_1,\dots,n_M}=\\
\tpd{\text{g}}{\partial}{\text{g}}+\sum_{i=1}^M (u_{n_i}^*, v_{n_i}^*)\partial
\begin{pmatrix}
	u_{n_i}\\
	v_{n_i}
\end{pmatrix}
\label{eqn:berry}.
\end{multline}
So the difference between the Berry phase of a state with quasiparticles and the ground state is simply the sum of ``Berry phase'' of the corresponding BdG wavefunctions. Since the Berry phase of ground state $\ket{\text{g}}$ can be eliminated by a global $\mathbb{U}(1)$ transformation, only the difference has physical meaning.

 According to \eqref{eqn:berry}, the relevant term to be evaluated is
\begin{multline}
	(u_{01}^*-iu_{02}^*, v_{01}^*-iv_{02}^*)\partial
\begin{pmatrix}
	u_{01}+iu_{02}\\
	v_{01}+iv_{02}
\end{pmatrix}
=\\
2\mathrm{Re}\left(u_1^*\partial u_1+u_2^*\partial u_2\right) 
+2i\mathrm{Re}(u_1^*\partial u_2-u_2^*\partial u_1),
\label{eqn:calc1}
\end{multline}
where we have made use of the Majorana condition $v=u^*$. The first term in \eqref{eqn:calc1} vanishes because $\int u^*\partial u$ must be purely imaginary. The second term has a non-vanishing contribution to the total Berry phase. However, due to the localized nature of zero-energy state, the overlap between $u_1$ and $u_2$ is exponentially small:
\begin{equation}
\int_0^T\di t\,\mathrm{Re}(u_1^*\partial_t u_2-u_2^*\partial_t u_1)\sim  e^{-|\vec{R}_1-\vec{R}_2|/\xi}.
\end{equation}
Therefore the Berry phase can be neglected in the limit of large separation $R$. This completes our discussion of non-Abelian statistics. The above calculation can be easily generalized to the case of many anyons.

\section{Time-dependent Bogoliubov-de Gennes Equation}
\label{sec:bdg}
In this section we derive the formalism to track down time-evolution of BCS condensate within mean-field theory.  For BCS superconductivity arising from interactions, the pairing order parameter has to be determined self-consistently, which makes the mathematical problem highly nonlinear. In the situations that we are interested in, it is not critical where the pairing comes from. In some systems that are believed to be experimentally accessible, e.g. semiconductor/superconductor heterostructure, superconductivity is induced by proximity effect~\cite{Fu_PRL08, Sau_PRL10} and there is no need to keep track of the self-consistency. We will take the perspective that order parameter is simply a external field in the Hamiltonian.

The time-dependent BdG equation~\cite{Andreev_JETP1964, Kuemmel_1969} has been widely used to describe dynamical phenomena in BCS superconductors. To be self-contained here we present a derivation of the time-dependent BdG equation highlighting its connection to quasiparticle operators. It can also be derived by methods of Heisenberg equation of motion or Green's function. Suppose we have a time-dependent BdG Hamiltonian ${H}_\text{BdG}(t)$. The unitary time-evolution of the many-body system is formally given by 
\begin{equation}
	\hat{U}(t)=\mathcal{T}\exp\left[ -i\int_0^t\di t'\, \hat{\mathcal{H}}_\text{BCS}(t') \right].
	\label{}
\end{equation}

To obtain an explicit form of $\hat{U}(t)$, we define the time-dependent Bogoliubov quasiparticle operator as
\begin{equation}
	\hat{\gamma}_n(t)=\hat{U}(t)\hat{\gamma}_n\hat{U}^\dag(t),
\end{equation}
where $\hat{\gamma}_n$ is the quasiparticle operator for $\hat{\mathcal{H}}_\text{BCS}(0)$ and the corresponding BdG wavefunction is $u_n(\vr), v_n(\vr)$. We adopt the normalization condition
\begin{equation}
  \int\di^2\vr\,|u_n(\vr)|^2+|v_n(\vr)|^2=1,
  \label{}
\end{equation}
which means $\{\hat{\gamma}_n,\hat{\gamma}_n^\dag\}=1$.

$\hat{\gamma}_n(t)$ by definition satisfies the following equation of motion
\begin{equation}
	i\odiff{\hat{\gamma}_n(t)}{t}=[\hat{\mathcal{H}}_\text{BCS}(t),\hat{\gamma}_n(t)].
	\label{}
\end{equation}

In fact, by direct calculation one can show that the equation of motion is solved by 
\begin{equation}
	\hat{\gamma}_n^\dag(t)=\int \di\vr\, \left[ u_n(\vr,t)\hat{\psi}(\vr) +v_n(\vr,t)\hat{\psi}(\vr)\right],
\end{equation}
where the wavefunction $u_n(\vr,t)$ and $v_n(\vr,t)$ are solutions of time-dependent BdG equation:
\begin{equation}
	i\odiff{}{t}
	\begin{pmatrix}
		u_n(\vr,t)\\
		v_n(\vr,t)
	\end{pmatrix}=\bm{H}_\text{BdG}(t)
	\begin{pmatrix}
		u_n(\vr,t)\\
		v_n(\vr,t)
	\end{pmatrix}
	\label{eqn:tdbdg}
\end{equation}
together with initial condition (another way of saying $\hat{\gamma}_n(0)=\hat{\gamma}_n$) 
\begin{equation}
	u_n(\vr,0)=u_n(\vr), v_n(\vr,0)=v_n(\vr).
\end{equation}
As long as the solutions of the time-dependent BdG equation \eqref{eqn:tdbdg} are obtained, we can construct the operators $\{\hat{\gamma}_n(t)\}$.

We now derive an explicit formula of $\hat{U}$ when the time-evolution is cyclic({\it i.e.}, $\hat{\mathcal{H}}_\text{BCS}(T)=\hat{\mathcal{H}}_\text{BCS}(0)$). In that case, it is always possible to express $\gamma_n(T)$ as a linear combination of $\hat{\gamma}_n\equiv\hat{\gamma}_n(0)$ and $\hat{\gamma}_n^\dag$. Since particle number is not conserved, it is more convenient to work with Majorana operators. We thus write $\hat{\gamma}_n=\hat{c}_{2n-1}+i\hat{c}_{2n}$ where $c_{m}$ are Majorana operators. Suppose the BdG matrix has totally $2N$ eigenvectors so $n=1,2\dots, N$. Assume that by solving time-dependent BdG equation we obtain the transformation of $c_m$ as follows:
\begin{equation}
	\hat{c}_{k}(T)=\sum_{l}\bm{V}_{kl}\hat{c}_l,
	\label{}
\end{equation}
where $\bm{V}\in \mathbb{SO}(2N)$ as required by unitarity and the conservation of fermion parity. The matrix $\bm{V}$ can be calculated once we know the BdG wavefunctions. Then we can write down an explicit expression of $\hat{U}(T)$ in terms of $\hat{c}_m$~\cite{Kitaev_Majorana}:
\begin{equation}
	\hat{U}(T)=\exp\left(\frac{1}{4} \sum_{mn} \bm{D}_{mn}\hat{c}_m\hat{c}_n \right),
	\label{UEXPR}
\end{equation}
where the matrix $\bm{D}$ is defined by the relation $e^{-\bm{D}}=\bm{V}$. Here $\bm{D}$ is necessarily a real, skew-symmetric matrix. A proof of this result is given in Appendix \ref{sec:app1}. We notice that the usefulness of \eqref{UEXPR} is actually not limited to cyclic evolution. In fact, \eqref{UEXPR} is purely an algebraic identity that shows any $\mathbb{SO}(2N)$ rotation of $2N$ Majorana operators can be implemented by a unitary transformation.

In the following we outline the method to solve the time-dependent BdG equation. To make connection with the previous discussion of Berry phase, we work in the ``instantaneous'' eigenbasis of time-dependent Hamiltonian. At each moment $t$, the BdG Hamiltonian $\bm{H}_\text{BdG}(t)$ can be diagonalized yielding a set of orthonormal eigenfunctions $\{\Psi_n(\vr,t)\}$. A remark is right in order: because of particle-hole symmetry, the spectrum of BdG Hamiltonian is symmetric with respect to zero energy and the quasiparticle corresponding to negative energy are really ``holes'' of positive energy states. However, at the level of solving BdG equation mathematically, both positive and negative energy eigenstates have to be retained to form a complete basis. The most general form of BdG wavefunction can be expanded as
\begin{equation}
	\Psi(\vr,t)=\sum_n c_n(t)\Psi_n(\vr,t).
	\label{}
\end{equation}

Plugging into time-dependent BdG equation, we obtain 
\begin{equation}
i\dot{c}_n + \sum_m \mathcal{M}_{nm}(t)c_m = E_n(t) c_n,
\label{eqn:bdg2}
\end{equation}
where $\mathcal{M}_{nm}=i\int\di^2\vr\Psi^\dag_n(\vr)\dot{\Psi}_m(\vr)$.

Assume that starting from initial condition $c_n(0)=\delta_{mn}$(roughly the quasiparticle is in the $\Psi_m$ state at $t=0$), we obtain the solutions of \eqref{eqn:bdg2} at $t=T$ denoted by $c^m_n(T)$. The transformation of basis states themselves is given by the matrix $\hat{B}$. Combining these two transformations we find
\begin{equation}
	\hat{\gamma}_n\rightarrow \sum_{kl}c^n_k \hat{B}_{kl}\hat{\gamma}_l,
	\label{}
\end{equation}
from which the linear transformation $\bm{V}$ can be directly read off. Then by taking the matrix log of $\hat{V}$ we can obtain the evolution operator. This is the procedure that we will use to solve the (cyclic) dynamics of BCS superconductors.

We will not be attempting to obtain the most general solution, since it depends heavily on the microscopic details. Instead, we focus on two major aspects of non-adiabaticity: (a) finite splitting of ground state degeneracy which only becomes appreciable when the braiding time is comparable to the ``tunneling'' time of Majorana fermions. (b) excited states outside the ground state subspace. 

Since these effects are highly non-universal, we have to specify how the braiding is performed. For example, when considering the braiding of non-Abelian vortices in two-dimensions, we assume that the trajectory of the two vortices is a circle and the braiding is performed at a constant speed. Mathematically, we choose $\vec{R}_1(t)$ and $\vec{R}_2(t)$ as  
\begin{equation}
  \begin{gathered}
    \vec{R}_1(t)=-\vec{R}_2(t)=R(\cos(\omega t+\theta_0), \sin(\omega t+\theta_0)).
  \end{gathered}
  \label{}
\end{equation}
Here we have defined $\omega=\frac{2\pi}{T}$ characterizing the energy scale corresponding to non-adiabaticity of the braiding itself. 

\subsection{Effect of tunneling splitting}
The derivation of transformation rule \eqref{eqn:ivanov} assumes that the two Majorana bound states have vanishing energies so there is no dynamical phase accumulated. The assumption is only true when tunneling splitting of zero-energy states is neglected. It has been established that the finite separation between anyons always leads to a non-zero splitting of zero-energy states~\cite{read_prb'00, Cheng_PRL09, Kraus_prl'09} although the splitting is exponentially suppressed due to the existence of bulk gap. As a result, the two ground states acquire different dynamical phases during the time-evolution. Here we take into account all the non-universal microscopic physics including dynamical phase induced by tunneling splitting and non-Abelian Berry phase.

In the framework of time-dependent BdG equation, the two basis states are $\Psi_{\pm}=\frac{1}{\sqrt{2}}(\Psi_{01}\pm i\Psi_{02})$. The energy splitting between two zero-energy states in vortices in a spinless $p_x+ip_y$ superconductor has been calculated in the limit of large separation~\cite{Cheng_PRL09, Cheng_PRB2010b}:
\begin{equation}
	E_+=-E_-\approx\sqrt{\frac{2}{\pi}}\Delta_0\frac{\cos(k_F R+\pi/4)}{\sqrt{k_FR}}e^{-R/\xi},
	\label{}
\end{equation}
with $\Delta_0$ being the amplitude of the bulk superconducting gap, $k_F$ the Fermi momentum and $\xi$ the coherence length. The exponential decay of splitting is universal for all non-Abelian topological phase and is in fact the manifestation of the topological protection.

The Berry matrix $\bm{M}$ can be evaluated and it takes the following form:
\begin{equation}
	\bm{M}=
\omega\begin{pmatrix}
	0 & \alpha\\
	\alpha^* & 0
\end{pmatrix},
	\label{}
\end{equation}
where $\alpha\equiv \alpha(R)$ is expressed as the overlap integral of the two bound state wavefunctions. The form of $\alpha$ is not important apart from the fact that $|\alpha|\sim e^{-R/\xi}$.

Since both $\alpha$ and $E_\pm$ are functions of $R$,  they are time-independent. We now have to solve essentially textbook problem of the Schr\"odinger equation of a spin $1/2$ in a magnetic field, the solution of which is well-known:
\begin{widetext}
\begin{equation}
	\begin{pmatrix}
		c_+(T)\\
		c_-(T)
	\end{pmatrix}=
	\begin{pmatrix}
		\cos\mathcal{E}T-\frac{i E_+}{\mathcal{E}}\sin\mathcal{E}T & \frac{i\omega\alpha}{\mathcal{E}}\sin\mathcal{E}T\\
		\frac{i\omega\alpha^*}{\mathcal{E}}\sin\mathcal{E}T & \cos\mathcal{E}T+\frac{i E_+}{\mathcal{E}}\sin\mathcal{E}T\\
	\end{pmatrix}
	\begin{pmatrix}
		c_+(0)\\
		c_-(0)
	\end{pmatrix},\:\:\mathcal{E}=\sqrt{E_+^2+\omega^2|\alpha|^2}.
	\label{}
\end{equation}
\end{widetext}

We can translate the results into transformation of Majorana operators:
\begin{equation}
	\begin{split}
		\hat{\gamma}_1\!&\rightarrow\!\Big(\cos\mathcal{E}T \!+\! \frac{i\omega\Re \alpha}{\mathcal{E}}\sin\mathcal{E}T\Big)	\hat{\gamma}_2\!-\!\frac{E_+\!-\!i\omega\Im \alpha}{\mathcal{E}}\sin\mathcal{E}T\hat{\gamma}_1\\
		\hat{\gamma}_2\!&\rightarrow\! - \frac{E_+\!+\!i\omega\Im \alpha}{\mathcal{E}}\sin\mathcal{E}T\hat{\gamma}_2\!-\! \Big(\cos\mathcal{E}T\!-\!\frac{i\omega\Re \alpha}{\mathcal{E}}\sin\mathcal{E}T\Big)\hat{\gamma}_1
	\end{split}.
	\label{}
\end{equation}

Because $E_+\sim \Delta_0 e^{-R/\xi}, |\alpha|\sim e^{-R/\xi}$ and $\omega\ll \Delta_0$, by order of magnitude we can safely assume $|E_+|\gg \omega|\alpha|$. In the limiting case $\omega\rightarrow 0$ we find
\begin{equation}
\begin{split}
	\hat{\gamma}_1\rightarrow \cos\mathcal{E}T \hat{\gamma}_2-\sin\mathcal{E}T\hat{\gamma}_1\\
\hat{\gamma}_2\rightarrow -\sin\mathcal{E}T \hat{\gamma}_2-\cos\mathcal{E}T\hat{\gamma}_1
\end{split},
	\label{}
\end{equation}
which can be compactly written as $\hat{\gamma}_i\rightarrow \hat{U}\hat{\gamma}_i\hat{U}^\dag$ where
\begin{equation}
	\hat{U}=\exp\left[ \left( \frac{\pi}{4}-\frac{\mathcal{E}T}{2} \right)\hat{\gamma}_2\hat{\gamma}_1\right].
	\label{}
\end{equation}

When we also take into account the terms containing $\omega$, the transformation matrix is no longer in $\mathbb{SO}(2)$. This simply means that we can no longer make the assumption that the system stays in the ground state subspace. 

\subsection{Effects of Excited Bound States}
The concept of quantum statistics is built upon the adiabatic theorem claiming that in adiabatic limit, quantum states evolve within the degenerate energy subspace. Going beyond adiabatic approximation, we need to consider processes that can cause transitions to states outside the subspace which violates the very fundamental assumption of adiabatic theorem.  In the case of the braiding of Majorana fermions in superconductors, there are always extended excited states in the spectrum which are separated from ground states by roughly the superconducting gap. In addition, there may be low-lying bound states within the bulk gap, such as the CdGM states in vortices. We call them subgap states. Extended states and subgap bound states apparently play different roles in the braiding of Majorana fermions. To single out their effects on the braiding we consider them separately and in this subsection we consider excited bound states first. Since the energy scale involved here is the superconducting gap, we neglect the exponentially small energy splitting whose effect has been considered in the previous subsection.

The BdG wavefunctions of excited bound states in each defect are denoted by $\Psi_{\lambda i}, i=1,2$ where $i$ labels the defects, with energy eigenvalues $\varepsilon_\lambda$. In Appendix \ref{app2} we show  the analytical forms of these wavefucntions for bound states in vortices known as CdGM bound states. If no other inhomogeneities are present, the wavefunctions are all functions of $\vr-\vec{R}_i$. Assuming $|\vec{R}_i-\vec{R}_j|\gg \xi$, we have approximately  $\pd{\Psi_{\lambda i}}{\Psi_{\lambda' j}}=\delta_{ij}$ up to exponentially small corrections. Therefore, the BdG equation decouples for the two defects since the tunneling amplitudes between them are all negligible. So it is sufficient to consider one defect and we will omit the defect label $i$ in the following. We write the solution to time-dependent BdG equation as
\begin{equation}
\Psi(t)\!=\!c_0(t)\Psi_0(t)+\sum_{\lambda}\left[c_\lambda(t)\Psi_\lambda(t)+\ol{c}_\lambda(t)\ol{\Psi}_\lambda(t)\right],
\end{equation}
where we have defined $\ol{\Psi}_\lambda$ as the particle-hole conjugate state of $\Psi_\lambda$, with energy eigenvalue $-\varepsilon_\lambda$.  We will focus on the minimal case where only one extra excited state is taken into account. Actually, in the case of bound states in vortices, due to the conservation of angular momentum, zero-energy state is only coupled to one excited bound state (See Appendix \ref{app2}) and to the leading order we can neglect the couplings of the zero-energy states to other excited states as well as those between excited states. The time-dependent BdG equation reduces to
\begin{equation}
\begin{split}
i\dot{c}_0&=-\beta c_\lambda+\beta^* \ol{c}_\lambda\\
i\dot{c}_\lambda&=(\varepsilon_\lambda-\alpha)c_\lambda-\beta^*c_0\\
i\dot{\ol{c}}_\lambda&=-(\varepsilon_\lambda+\alpha)\ol{c}_\lambda-\beta c_0
\end{split},
\label{redbdg1}
\end{equation}
where we have defined the components of the Berry matrix as $\beta_\lambda=i\pd{\Psi_0}{\dot{\Psi}_\lambda}, \alpha_\lambda=i\pd{\Psi_\lambda}{\dot{\Psi}_\lambda}$.  Expressions of $\beta$ and $\alpha$ can be found in Appendix \ref{app2}. In the following we suppress the subscript $\lambda$ and shift the energy of excited level to eliminate $\alpha_\lambda$: $\varepsilon_\lambda\rightarrow \varepsilon_\lambda+\alpha_\lambda$. Solving the BdG equation \eqref{redbdg1}, we find
\begin{equation}
	\begin{split}
      \hat{\gamma}&\rightarrow \left( 1-\frac{4\beta^2}{E^2}\sin^2\frac{Et}{2} \right)\hat{\gamma}+\frac{2\sqrt{2}\beta\varepsilon\sin^2\frac{Et}{2}}{E^2}\hat{\xi}-\frac{\sqrt{2}\beta\sin Et}{E}\hat{\eta}\\
      \hat{\xi}&\rightarrow \frac{2\sqrt{2}\varepsilon\beta\sin^2\frac{Et}{2}}{E^2}\hat{\gamma} +\left(\cos Et+\frac{4\beta^2\sin^2\frac{Et}{2}}{E^2}\right)\hat{\xi} + \frac{\varepsilon\sin Et}{E}\hat{\eta}\\
      \hat{\eta}&\rightarrow \frac{\sqrt{2}\beta \sin Et}{E}\hat{\gamma}-\frac{\varepsilon\sin Et}{E}\hat{\xi}+\cos Et\,\hat{\eta}
	\end{split}
	\label{eqn:res1}
\end{equation}
which should be followed up by the basis transformation $\hat{B}$. Here $E=\sqrt{\varepsilon^2+4\beta^2}$.

By using \eqref{UEXPR} we can work out explicitly how the ground state wavefunctions transform. Physically, the non-adiabatic process causes transitions of quasiparticles residing on the zero-energy level to the excited levels.  
Superficially these transitions to excited states significantly affect the non-Abelian statistics, since the parity of fermion occupation in the ground state subspaces is changed as well as the quantum entanglement between various ground states~\cite{Stern_PRB'04, Moller_PRB2011}. This can also be directly seen from \eqref{eqn:res1} since starting from $\ket{\text{g}}$ the final state is a superposition of $\ket{\text{g}}$ and $\hat{d}_0^\dag \hat{d}_\lambda^\dag\ket{\text{g}}$. So we might suspect that errors are introduced to the gate operations. However, noticing that the excited states are still localized, the parity of the total fermion occupation in the ground state subspace and local excited states are well conserved. This observation allows for a redefinition of the Majorana operators to properly account for the fermion occupation in local excited states, as done in [\onlinecite{Akhmerov_PRB2010}]. We therefore have generalized conservation of ``local'' fermion parity
\begin{equation}
	\hat{\mathcal{P}}_{12}=i\hat{\gamma}_1\hat{\gamma}_2 \prod_{i=1,2}(1-2\hat{d}^\dag_i \hat{d}_i).
	\label{eqn:fermionparity}
\end{equation}
shared by defects 1 and 2. It is not difficult to see that these generalized fermion parity operators(thus their expectation values) transform exactly according to Ivanov's rule, thus naturally extending non-Abelian statistics to an enlarged Hilbert space. From the perspective of measurement, to probe the status of a topological qubit it is necessary and sufficient to measure the fermion parity as defined in \eqref{eqn:fermionparity}. It is practically impossible(and unnecessary) to distinguish between the fermion occupations in ground state subspace and excited states as long as they are both localized and can be considered as a composite qubit.

\subsection{Effects of Excited Extended States}
We next consider the effect of excited states that are extended in space. Usually such states form a continuum. The scenario considered here may not be very relevant to Majorana fermions in 2D $p_x+ip_y$ superconductor since bound states in vortices often dominate at low energy. However, for Majorana fermions in one-dimensional systems, zero-energy state is the only subgap state and coupling between zero-energy state and continuum of excited states may become important. 

Again we write time-dependent BdG equation in the instantaneous eigenbasis
\begin{equation}
\begin{gathered}
	i\dot{c}_{0,1}= -\sum_\lambda\left(\beta_{1\lambda}  c_{\lambda}- \beta_{1\lambda}^*\ol{c}_{\lambda}\right)\\
	i\dot{c}_{0,2}= -\sum_\lambda\left(\beta_{2\lambda}c_{\lambda}- \beta_{2\lambda}^*\ol{c}_{\lambda}\right)\\
	i\dot{c_{\lambda}}=(\varepsilon_\lambda-\alpha_\lambda)c_{\lambda}-\beta_{1\lambda}^* c_{0,1}-\beta_{2\lambda}^* c_{0,2}\\
	i\dot{\ol{c}_\lambda}=-(\varepsilon_\lambda-\alpha_\lambda)\ol{c}_{\lambda}+\beta_{1\lambda} c_{0,1} +\beta_{2\lambda} c_{0,2}
\end{gathered}.
\label{eqn:deq3}
\end{equation}
Here we still use $\lambda$ to label the excited states. As in the case of bound states, we ignore the coupling between excited states since these only contribute higher order terms to the dynamics of zero-energy states. In another word, we treat each excited state individually and in the end their contributions are summed up.

To proceed we need to determine matrix elements $\beta_{1\lambda}$ and $\beta_{2\lambda}$. We will consider the spinless one-dimensional p-wave superconductor as an example. The zero-energy states localize at two ends of the 1D system which lie in the interval $[0, L]$. We also assume reflection symmetry with respect to $x=L/2$. Without worrying about the tunneling splitting, we can consider the two ends near $x=0$ and $x=L$ independently. Then we make use of the fact that BdG equations near $x=0$ and $x=L$ are related by a combined coordinate and gauge transformation $x\rightarrow L-x$, $\Delta(x)\rightarrow -\Delta(L-x)$. Therefore, the bound state and the local part of the extended states near $x=0$ and $x=L$ are related by gauge transformations. Based on these considerations, we should have $\beta_{1\lambda}=\beta_{2\lambda}\equiv \beta_\lambda$, up to exponentially small corrections.  Similar argument also applies to vortices in 2D $p_x+ip_y$ superconductors.

As mentioned above, we will make the approximation that each excited state can be treated independently. So we consider the effect of one of the excited states first and omit the label $\lambda$ temporarily. Again we write $\hat{d}=\frac{1}{\sqrt{2}}(\hat{\xi}+i\hat{\eta})$. Without loss of generality we also assume that $\beta$ is real. We find from the solution of time-dependent BdG equation that
\begin{widetext}
\begin{equation}
  \begin{split}
	  \hat{\gamma}_1&\rightarrow\frac{4\beta^2}{{E}^2}\sin^2\frac{{E}t}{2}\hat{\gamma}_1+ \left(1-\frac{4\beta^2}{{E}^2}\sin^2\frac{{E}t}{2}\right)\hat{\gamma}_2+\frac{2\sqrt{2}\beta\varepsilon\sin^2\frac{Et}{2}}{E^2}\hat{\xi}+\frac{\sqrt{2}\beta\sin Et}{E}\hat{\eta} \\
	  \hat{\gamma}_2&\rightarrow -\left(1-\frac{4\beta^2}{{E}^2}\sin^2\frac{{E}t}{2}\right)\hat{\gamma}_1-\frac{4\beta^2}{{E}^2}\sin^2\frac{{E}t}{2}\hat{\gamma}_2+\frac{2\sqrt{2}\beta\varepsilon\sin^2\frac{Et}{2}}{E^2}\hat{\xi}+\frac{\sqrt{2}\beta\sin Et}{E}\hat{\eta} \\
	  \hat{\xi}&\rightarrow \frac{2\sqrt{2}\beta\varepsilon\sin^2 \frac{Et}{2}}{E^2}(-\hat{\gamma}_1+\hat{\gamma}_2)+\left( \cos Et+\frac{8\beta^2\sin^2\frac{Et}{2}}{E^2} \right)\hat{\xi}-\frac{\sqrt{2}\varepsilon\sin Et}{E}\hat{\eta}\\
	  \hat{\eta}&\rightarrow \frac{2\sqrt{2}\beta\sin^2 \frac{Et}{2}}{E}(-\hat{\gamma}_1+\hat{\gamma}_2)+\frac{\varepsilon \sin Et}{E}\hat{\xi}+\cos Et\,\hat{\eta}
  \end{split}.
  \label{eqn:res3}
\end{equation}
\end{widetext}
Here again $E=\sqrt{\varepsilon^2+4\beta^2}$.

At first glance the physics here is very similar to what has been discussed for local bound states: non-adiabatic transitions cause changes of fermion parity in the ground state subspace. The crucial difference between local bound states and a continuum of extended states is that, in the former case, local fermion parity is still conserved as long as we count fermion occupation in the excited states while in the latter, it is impossible to keep track of the number of fermions leaking into the continuum so the notion of local fermion parity breaks down. These non-adiabatic effects may pose additional constraints on manufacturing of topological qubits. Let's consider to what extent the braiding statistics is affected. A useful quantity to look at here is the expectation value of the fermion parity operator in the ground state subspace, namely $\langle\hat{\mathcal{P}}_0\rangle=\langle i\hat{\gamma}_1\hat{\gamma}_2\rangle $.  

Suppose at $t=0$ we start from the ground state $\ket{\text{g}}$ with even fermion parity $\tpd{\text{g}}{\hat{\mathcal{P}}_0}{\text{g}}=1$ and the excited level is unoccupied, too. After the braiding at time $T$ the expectation value of $\hat{\mathcal{P}}_0$ becomes
\begin{equation}
	\langle{\hat{\mathcal{P}}_0(T)}\rangle=1\!-\!\frac{8\beta^2}{E^2}\sin^2\frac{ET}{2}\left(1\!+\!\frac{2\varepsilon}{E}\sin^2\frac{ET}{2}\right).
  \label{eqn:parity2}
\end{equation}
where $\hat{\mathcal{P}}_0(T)=\hat{U}^\dag(T)\hat{\mathcal{P}}_0\hat{U}(T)$. This confirms that fermion parity is not conserved anymore. For $|\beta|\ll\varepsilon$, the coupling to excited state can be understood as a small perturbation. $\langle\hat{\mathcal{P}}_0\rangle$ only slightly deviates from the non-perturbed value. In the opposite limit $|\beta|\gg \varepsilon$, $\langle\hat{\mathcal{P}}_0\rangle$ can oscillate between $1$ and $-1$ so basically fermion parity is no longer well-defined.

Now we can sum up the contributions from each excited state and \eqref{eqn:parity2} is replaced by:
\begin{equation}
	\langle{\hat{\mathcal{P}}_0(T)}\rangle=1\!-\!\sum_\lambda\frac{8|\beta_\lambda|^2}{E_\lambda^2}\sin^2\frac{E_\lambda T}{2}\left(1\!+\!\frac{2\varepsilon_\lambda}{E_\lambda}\sin^2\frac{E_\lambda T}{2}\right).
  \label{eqn:parity3}
\end{equation}

\section{Discussion and Conclusion}
In conclusion, we have considered the braiding of non-Abelian anyons as a dynamical process and calculated the corrections to non-Abelian evolutions due to non-adiabatic effects. We discuss several sources of non-adiabaticity: first of all, tunneling between non-Abelian anyons results in splitting of the degenerate ground states. The Abelian dynamical phase accumulated in the process of braiding modifies Ivanov's rule of non-Abelian statistics. Since the bulk of the superconductor is fully gapped such corrections are exponentially small. In the context of TQC such deviations from Ivanov's rule are sources of errors in single-qubit quantum gates.

Second, we consider dynamical transitions of Majorana fermions in the zero-energy ground states to excited states. The effects of such non-adiabatic transitions strongly rely on whether these excited states are bound states with discrete spectrum and localized at the same positions with the Majorana bound states, or they extend through the whole bulk and form a continuum. Generally speaking, non-adiabatic transitions mix the zero-energy ground states with other excited states and it is questionable whether the quantum entanglement crucial to non-Abelian statistics is still preserved. In the former case where excited states are localized, we are still able to define conserved fermion parity stored in these low-energy bound states. Non-Abelian statistics can be generalized once we enlarge the Hilbert space to include all local bound states. In the latter case, the situation is dramatically different because the notion of fermion parity in the low-energy Hilbert space no longer makes sense once extended states above the bulk gap are involved. We characterize the loss of fermion parity in such non-adiabatic transitions by the expectation value of the ``local fermion parity'' operator.  This can be viewed as the dissipation of topological qubit resulting from couplings to a continuum of fermionic states. We have thus quantified the expectation that a zero-energy Majorana mode will decay if it is put in contact with a continuum of fermionic states({\it e.g.}, electrons).

Although the underlying technological motivation for topological quantum computation is that quantum error correction against continuous decoherence is unnecessary as a matter of principle in topological systems since decoherence due to local coupling to the environment is eliminated, other errors, such as non-adiabaticity considered in this work, would invariably occur in all quantum systems in the presence of time-dependent quantum gate operations.  In addition, braiding is the cornerstone of the strange quantum statistical properties which distinguish non-Abelian anyons from ordinary fermions and bosons.  Our work, involving the non-adiabatic corrections to anyonic braiding, is therefore relevant to all current considerations in the subject of Ising anyons whether it is in the context of the observation of the non-Abelian statistics or the implementation of topological quantum computation. In particular, non-adiabaticity in the Majorana braiding in the specific context of non-Abelian topological superconductors as discussed in this paper, may be relevant to various recently proposed Majorana interferometry experiments involving vortices in 2D~\cite{Fu_PRL2009, Akhmerov_PRL2009, Sau_arxiv1004, Grosfeld_PRB2010, Grosfeld_arxiv1012}.

We now speculate about the physical sources of non-adiabaticity of braidings. Throughout our work we have focused on the intrinsic non-adiabaticity originating from the fact that braidings are done during a finite interval of time. For such effects to be appreciable, the time-scale of braidings has to be comparable to $\Delta t\sim \frac{\hbar}{\Delta E}$ where $\Delta E$ is the energy gap protecting the anyonic Majorana modes. For the corrections from the tunneling of Majorana fermions, $\Delta t$ is exponentially large so it is usually legitimate to neglect the tunneling effect. On the other hand, if we are interested in corrections from states above the gap, then the relevant time scale is $\hbar/\Delta$. Using \eqref{eqn:parity3} one can estimate the non-adiabatic error rate in performing gate operations. 

These considerations also apply to other types of non-adiabatic perturbations as long as they have non-zero matrix elements between zero-energy states and excited states. In particularly disorder scattering, which is unavoidable in solid state systems, can be a source of non-adiabaticity~\cite{Moller_PRB2011}. As the non-Abelian anyons are moving, the disorder potential seen by the anyons changes randomly with time so it can be modeled as a time-dependent noise term in the Hamiltonian which may cause dephasing. Other possible perturbations in solid state systems include collective excitations, such as phonons and plasmons(or phase fluctuations). We leave the investigation of these effect for future work. The formalism developed in the current work can, in principle, be used to study these dephasing errors.

\section{Acknowledgements}

\appendix
We thank Xin Wang for a critical reading of the manuscript. This work is supported by DARPA-QuEST, JQI-NSF-PFC, and Microsoft Station Q.
\section{Proof of Eq. \eqref{UEXPR}}
\label{sec:app1}
This result already appears in Ref. [\onlinecite{Kitaev_Majorana}]. To be self-contained we give a proof here. Assume that $\bm{D}$ is not singular. Then there exists a real orthogonal matrix $\bm{O}$ such that  
\begin{equation}
	\bm{O}\bm{D}\bm{O}^T=
	\begin{pmatrix}
	0 & \theta_1 & & & \\
	-\theta_1 & 0 & & &\\
	 & &\ddots & &\\
	& & & 0 & \theta_N\\
	& & & -\theta_N & 0
\end{pmatrix}\equiv \bm{\theta},
	\label{}
\end{equation}
where $\theta_n>0$. We can find a new set of Majorana operators
\begin{equation}
	\begin{pmatrix}
		\hat{b}_1\\
		\hat{b}_1'\\
		\vdots\\
		\hat{b}_N\\
		\hat{b}_N'
	\end{pmatrix}=\bm{O}
	\begin{pmatrix}
		\hat{c}_1\\
		\hat{c}_2\\
		\vdots\\
		\hat{c}_{2N-1}\\
		\hat{c}_{2N}
	\end{pmatrix},
	\label{}
\end{equation}
that allow us to write $\hat{c}^T \hat{B} \hat{c}$ in a canonical form:
\begin{equation}
	\sum_{kl}c_k \bm{D}_{kl} c_l=2\sum_m \theta_m \hat{b}_m\hat{b}_m'.
	\label{}
\end{equation}
Therefore we have
\begin{equation}
	\hat{U}=\exp\left(\frac{1}{4}\hat{\bm{c}}^T \bm{D}\hat{\bm{c}} \right)=\prod_m\left(\cos\frac{\theta_m}{2}+\sin\frac{\theta_m}{2}\hat{b}_m\hat{b}_m'\right).
	\label{}
\end{equation}

The transformation of $\hat{b}_n, \hat{b}_n'$ under $\hat{U}$ can be easily worked out:
\begin{equation}
	\begin{gathered}
	\hat{U}\hat{b}_n\hat{U}^\dag=\cos\theta_n b_n-\sin\theta_n b_n'\\
\hat{U}\hat{b}_n'\hat{U}^\dag=\sin \theta_n b_n+\cos \theta_n b_n'
\end{gathered},
	\label{}
\end{equation}

We write it in matrix form:
\begin{equation}
	\hat{U}\hat{b}\hat{U}^\dag=
\begin{pmatrix}
	\cos\theta_1 & -\sin\theta_1 & & & \\
	\sin\theta_1 & \cos \theta_1 & & &\\
	 & &\ddots & &\\
	& & & \cos \theta_N & -\sin\theta_N\\
	& & & \sin\theta_N & \cos \theta_N
\end{pmatrix}\hat{b}.
	\label{}
\end{equation}

Notice that
\begin{equation}
\begin{pmatrix}
	\cos\theta_n & -\sin\theta_n\\
	\sin\theta_n & \cos \theta_n\\
\end{pmatrix}=\exp\left[ 
-\begin{pmatrix}
	0 & \theta_n\\
	-\theta_n & 0
\end{pmatrix}
\right],
\end{equation}
so the big matrix is essentially $e^{-\hat{\theta}}$. Now the rest of the proof is rather straightforward:
\begin{equation}
	\hat{U}\hat{\bm{c}}\hat{U}^\dag =\bm{O}^T e^{-\hat{\theta}} \hat{\bm{b}}=\bm{O}^T e^{-\bm{\theta}}\bm{O}\hat{\bm{c}}=e^{-\bm{D}}\hat{\bm{c}}.
	\label{}
\end{equation}
Thus we have $e^{-\bm{D}}=\bm{V}$.

Formula \eqref{UEXPR} gives the explicit form of unitary transformation corresponding to a cyclic time evolution. However, analytically finding matrix $\bm{D}$ as the logarithm of the matrix $\bm{V}$ is still quite difficult. In fact, for the purpose of calculating the operator expectation values, it is not necessary to know explicitly $\bm{D}$. To illustrate this point, consider the ground state expectation value of $\hat{A}=\hat{c}_{i_1}\hat{c}_{i_2}\cdots\hat{c}_{i_l}$ before and after braiding. We denote the starting ground state by $\ket{\psi}$. Then the expectation value on the transformed ground state after braiding is given by
\begin{equation}
	\tpd{\psi}{\hat{U}^\dag \hat{A} \hat{U}}{\psi}=\tpd{\psi}{\hat{U}^\dag \hat{c}_{i_1} \hat{U}\cdot\hat{U}^\dag \hat{c}_{i_2} \hat{U}\cdots\hat{U}^\dag \hat{c}_{i_l} \hat{U}}{\psi},
	\label{}
\end{equation}
where $\hat{U}^\dag \hat{\bm{c}} \hat{U}$ can be easily evaluated:
\begin{equation}
	\hat{U}^\dag \hat{\bm{c}} \hat{U}=e^{\bm{D}}\hat{\bm{c}}=\bm{V}^T\hat{\bm{c}}.
	\label{}
\end{equation}
\section{Bound states in vortices}
\label{app2}
In this appendix we present the analytical solutions of CdGM bound states~\cite{Caroli_PL'64} in spinless $p_x+ip_y$ superconductors~\cite{Kopnin_PRB'91,Mizushima_PRA10, Moller_PRB2011} in the weaking-coupling limit($\Delta\ll E_F$). We model the $p_x+ip_y$ pairing by the following gap operator:
\begin{equation}
	\hat{\Delta}(\vr)=-\frac{1}{k_F}\{ \Delta(\vr),\partial_x+i\partial_y\}.
	\label{}
\end{equation}

A vorticity $1$ vortex sitting at the origin is represented by the following order parameter 
\begin{equation}
	\Delta(\vr)=\Delta(r)e^{i\theta},
	\label{}
\end{equation}
where $(r,\theta)$ is the polar coordinate of 2D plane.

The (unnormalized) wave function of bound states are given by
\begin{equation}
\begin{pmatrix}
u_l(\vr)\\
v_l(\vr)
\end{pmatrix}=
e^{il\theta} 
\begin{pmatrix}
f_l(k_+ r)\\
f_{l-2}(k_-r)e^{-2i\theta}
\end{pmatrix}.
\label{}
\end{equation}
Here $f_l(k_\pm r)$ is defined to be
\begin{equation}
  f_l(k_\pm r)=J_l(k_\pm r) e^{-\frac{1}{v_F}\int_0^r\di r'\Delta(r')},
\label{}
\end{equation}
where $k_\pm = k_\mu\pm m\varepsilon_l/k_\mu$ and the corresponding energy levels
\begin{equation}
\varepsilon_l=-(l-1)\omega_0, \:\omega_0\sim \frac{\Delta_0^2}{\varepsilon_F}.
	\label{}
\end{equation}

In weak-coupling limit we neglect the difference between $\mu$ and Fermi energy $\varepsilon_F$ so $k_\pm \approx k_F$.

Zero energy state corresponds to $l=1$. Written out explicitly, it has the form
\begin{equation}
	\Psi_0(\vr)=
\begin{pmatrix}
	ie^{i\theta}\\
-ie^{-i\theta}
\end{pmatrix}
J_1(k_F r) e^{-\frac{1}{v_F}\int_0^r\di r'\Delta(r') }.
\label{}
\end{equation}
Notice that we have multiplied the wavefunction by $i$ to satisfy the Majorana condition: $v_{l=1}=u^*_{l=1}$.
The first excited state has $l=0$:
\begin{equation}
\begin{pmatrix}
u_{l=0}(\vr)\\
v_{l=0}(\vr)
\end{pmatrix}=
\begin{pmatrix}
J_0(k_F r)\\
J_{2}(k_Fr)e^{-2i\theta}
\end{pmatrix}
 e^{-\frac{1}{v_F}\int_0^r\di r'\Delta(r') }.
\label{}
\end{equation}

Having all these analytical expressions, we can calculate the Berry connections. We skip the algebra and the result is given by
\begin{equation}
  i\pd{\Psi_{l=1,i}}{\dot{\Psi}_{l,i}}=-i\pi A[(v_x-iv_y)\delta_l-(v_x+iv_y)\delta_{l-2}],
  \label{}
\end{equation}
where $\vec{v}=\dot{\vec{R}}$ and constant $A$ is the following integral:
\begin{multline}
A=\int_0^\infty r\di r\,f_1(k_Fr)\partial_rf_0(k_Fr)-\\
\int_0^\infty r\di r\,f_{1}(k_Fr)\left(\partial_r+\frac{2}{r}\right)f_{2}(k_F r).
\end{multline}


\end{document}